\documentclass[preprint,aps,superscriptaddress]{revtex4}

\usepackage{graphicx}
\usepackage{subfigure}
\usepackage{epsfig}
\usepackage{xcolor}
\usepackage{dcolumn}
\usepackage{bm}
\usepackage{ulem}
\usepackage{color}
\usepackage{amsmath}
\usepackage{amsfonts}
\usepackage{amssymb}

\usepackage{epstopdf}

\begin{document}

\title{Second generation Dirac cones and inversion symmetry breaking induced gaps in graphene/h-BN}

\author{Eryin Wang}
\altaffiliation{These authors contribute equally to this work.}
\affiliation{State Key Laboratory of Low Dimensional Quantum Physics and Department of Physics, Tsinghua University, Beijing 100084, China}

\author{Xiaobo Lu}
\altaffiliation{These authors contribute equally to this work.}
\affiliation{Beijing National Laboratory for Condensed Matter Physics and Institute of Physics, Chinese Academy of Sciences, Beijing 100190, China}

\author{Shijie Ding}
\affiliation{State Key Laboratory of Low Dimensional Quantum Physics and Department of Physics, Tsinghua University, Beijing 100084, China}

\author{Wei Yao}
\affiliation{State Key Laboratory of Low Dimensional Quantum Physics and Department of Physics, Tsinghua University, Beijing 100084, China}

\author{Mingzhe Yan}
\affiliation{State Key Laboratory of Low Dimensional Quantum Physics and Department of Physics, Tsinghua University, Beijing 100084, China}

\author{Guoliang Wan}
\affiliation{State Key Laboratory of Low Dimensional Quantum Physics and Department of Physics, Tsinghua University, Beijing 100084, China}

\author{Ke Deng}
\affiliation{State Key Laboratory of Low Dimensional Quantum Physics and Department of Physics, Tsinghua University, Beijing 100084, China}

\author{Shuopei Wang}
\affiliation{Beijing National Laboratory for Condensed Matter Physics and Institute of Physics, Chinese Academy of Sciences, Beijing 100190, China}

\author{Guorui Chen}
\affiliation{State Key Laboratory of Surface Physics and Department of Physics, Fudan University, Shanghai 200433, China}

\author{Liguo Ma}
\affiliation{State Key Laboratory of Surface Physics and Department of Physics, Fudan University, Shanghai 200433, China}

\author{Jeil Jung}
\affiliation{Department of Physics, University of Seoul, Seoul 02504, Korea}

\author{Alexei V. Fedorov}
\affiliation{Advanced Light Source, Lawrence Berkeley National Laboratory, Berkeley, CA 94720, USA}

\author{Yuanbo Zhang}
\affiliation{State Key Laboratory of Surface Physics and Department of Physics, Fudan University, Shanghai 200433, China}

\author{Guangyu Zhang}
\affiliation{Beijing National Laboratory for Condensed Matter Physics and Institute of Physics, Chinese Academy of Sciences, Beijing 100190, China}
\affiliation{Collaborative Innovation Center of Quantum Matter, Beijing, P.R. China}

\author{Shuyun Zhou}
\altaffiliation{Correspondence should be sent to syzhou@mail.tsinghua.edu.cn}
\affiliation{State Key Laboratory of Low Dimensional Quantum Physics and Department of Physics, Tsinghua University, Beijing 100084, China}
\affiliation{Collaborative Innovation Center of Quantum Matter, Beijing, P.R. China}

\date{\today}

\begin{abstract}

{\bf  Graphene/h-BN has emerged as a model van der Waals heterostructure \cite{GeimVDW}, and the band structure engineering by the superlattice potential has led to various novel quantum phenomena including the self-similar Hofstadter butterfly states \cite{GeimNature13, KimNature13, GBNgap,GeimNaturePhys2014}. Although newly generated second generation Dirac cones (SDCs) are believed to be crucial for understanding such intriguing phenomena, so far fundamental knowledge of SDCs in such heterostructure, e.g. locations and dispersion of SDCs, the effect of inversion symmetry breaking on the gap opening, still remains highly debated due to the lack of direct experimental results. Here we report first direct experimental results on the dispersion of SDCs in 0$^\circ$ aligned graphene/h-BN heterostructure using angle-resolved photoemission spectroscopy. Our data reveal unambiguously SDCs  at the corners of the superlattice Brillouin zone, and at only one of the two superlattice valleys. Moreover, gaps of $\approx$ 100 meV and $\approx$ 160 meV are observed at the SDCs and the original graphene Dirac cone respectively. Our work highlights the important role of a strong inversion symmetry breaking perturbation potential in the physics of graphene/h-BN, and fills critical knowledge gaps in the band structure engineering of Dirac fermions by a superlattice potential.}

\end{abstract}

\maketitle

Hexagonal boron nitride (h-BN) shares similar honeycomb lattice structure to graphene, yet its lattice is stretched by 1.8$\%$. Moreover, the breaking of the inversion symmetry by distinct boron and nitrogen sublattices  leads to a large band gap (5.97 eV) in the $\pi$ band, which is in sharp contrast to the gapless Dirac cones in graphene. By stacking graphene atop h-BN to form a van der Waals heterostructure  \cite{GeimVDW}, graphene/h-BN not only exhibits greatly improved properties for device applications, such as reduced ripples, suppressed charge inhomogeneities and higher mobility \cite{HoneNano2010, LeRoyNatureMater11}, but also provides unique opportunities for band structure engineering of Dirac fermions by a periodic potential \cite{LouieNaturePhys08, LouiePRL08}.  The superlattice potential induced by the lattice mismatch and crystal orientation can significantly modify the electronic properties of graphene and lead to various novel quantum phenomena, e.g. emergence of second generation Dirac cones (SDCs) which are crucial for the realization of Hofstadter butterfly states under applied magnetic field \cite{GeimNature13, KimNature13, GBNgap,GeimNaturePhys2014}, renormalization of the Fermi velocity \cite{ LouieNaturePhys08, FalkoMiniband, JvdBPRB12, Guinea},  gap opening at the Dirac point \cite{DFTgap, GBNgap,DGGPRL13, GeimSci14,NovoselovNphys2014}, topological currents \cite{GeimSci14} and gate-dependent pseudospin mixing \cite{FWangGate}.   Hence, understanding the effects of superlattice potential on the band structure of graphene is crucial for advancing its device applications, and for gaining new knowledge about the fundamental physics of Dirac fermions in a periodic potential.

Previously, the existence of SDCs has been deduced from scanning tunneling spectroscopy, resistivity and capacitance measurements \cite{LeRoyNaturePhys12,GeimNature13, ZhangGY,GeimNaturePhys2014}. However, such measurements are not capable of mapping out the electronic dispersion with momentum-resolved information, and the lack of direct experimental results has led to ambiguous and even conflicting results about the electronic spectra of SDCs and the existence of gaps. Although various theoretical models have been proposed \cite{JvdBPRB12, FalkoMiniband, JungPRB14, MoonPRB14, FasolinoPRL, JeilNcomm}, the  locations and dispersions of SDCs strongly depend on the parameters used to describe the inversion-symmetric and inversion-asymmetric superlattice potential modulation \cite{FalkoMiniband}.  Different choices of inversion-symmetric perturbation could result in either isolated or overlapping SDCs \cite{FalkoMiniband}, and the locations of SDCs could change from the edges of the superlattice Brillouin zone (SBZ) \cite{LouiePRL08,FalkoMiniband} to the corners \cite{JvdBPRB12,MoonPRB14,FalkoMiniband}. The inversion-asymmetric perturbation potential  can strongly affect the gap opening at the Dirac point. The gap opening in graphene/h-BN is a highly debated issue, with some theoretical and experimental studies arguing for its existence \cite{GBNgap, ZhangGY, DGGPRL13, LiZQ, MoonPRB14, Bokdam, Guinea, JeilNcomm} while others ruling it out \cite{LouieNaturePhys08, HoneNano2010, LeRoyNatureMater11, GeimNature13, JvdBPRB12}. Such knowledge gaps in understanding the fundamental electronic structure of graphene/h-BN call for a careful examination of the electronic structure by angle-resolved photoemission spectroscopy (ARPES) which can map out the dispersions of original Dirac cone and SDCs with both energy- and momentum- resolution and detect the gap opening directly if there is any.

ARPES studies of graphene/h-BN heterostructure had been challenging for several reasons. Firstly, the size of the heterostructure prepared by transferring graphene atop the h-BN substrate was typically a few micrometers ($\mu$m), much smaller than the typical ARPES beam size of 50-100 $\mu$m.  Secondly, because of the large Moir$\acute{e}$ periodicity ($\approx$ 14 nm, 56 times of graphene' s lattice constant for 0$^\circ$ aligned graphene/h-BN), the separation between the original Dirac cone of graphene and the cloned Dirac cones is extremely small, on the order of the reciprocal superlattice vector $G_s\approx$ 0.05  $\AA^{-1}$. Resolving band dispersions within such a small momentum space requires extremely high quality samples with sharp spectral features. Recently, high quality 0$^\circ$ aligned graphene/h-BN heterostructures with large size extending a few hundred $\mu$m have been successfully grown directly by remote plasma-enhanced chemical vapor deposition (R-PECVD)  \cite{ZhangGY}, and this has made our ARPES studies possible.

\begin{figure*}
\includegraphics[width=16.5 cm] {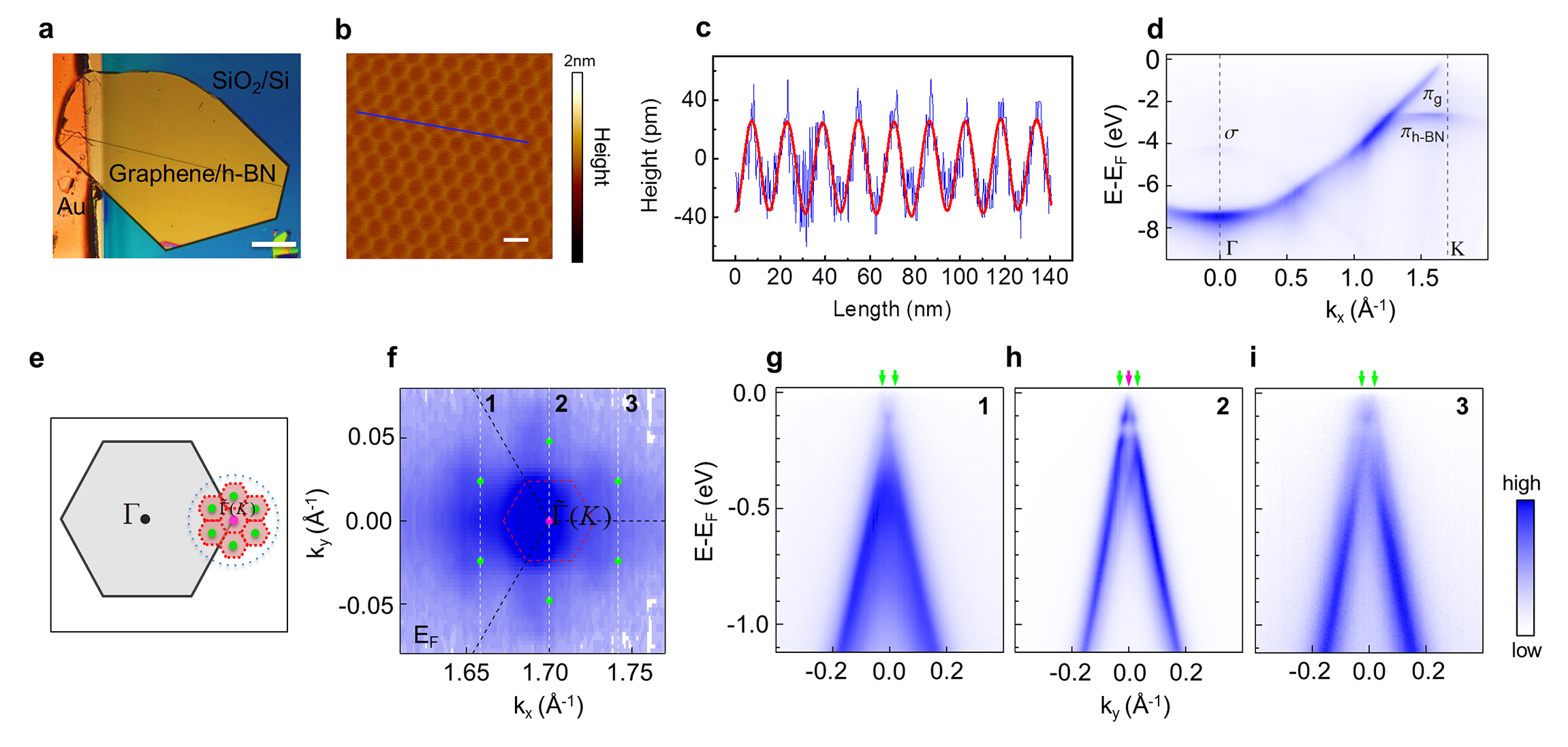}
\caption{{\bf Six cloned first generation Dirac cones from the superlattice potential.} (a) Optical micrograph image of a typical sample measured. The white scale bar is 100 $\mu m$. (b) Raw image of moir$\acute{e}$ pattern formed in Graphene/h-BN from atomic force microscopic taken in tapping mode in ambient atmosphere. The white scale bar is 20 nm. (c) Height profile along the blue line in(b). The raw data and the data after high-pass-filtered inverse fast Fourier transformation are shown in blue and red color respectively. (d) The dispersion along $\Gamma$-K direction to show the $\sigma$ and $\pi$ bands from graphene and h-BN. (e) Schematic drawing of Brillouin zones of graphene (grey hexagon) and SBZs (red dotted hexagons).  The $\Gamma$ (black circles), $\tilde{\Gamma}$ (K) (pink circles) points of graphene brillouin zone and the six nearest SBZ centers (green circles) are shown. (f) Constant energy map at E$_F$. Replicas are observed around the six nearest SBZ centers (green circles). The brillouin zones of graphene and superlattice are indicated by black and red broken lines respectively.  (g-i) Dispersions along cuts 1, 2, 3 indicated in (f). The tops of original Dirac cone and replicas are indicated by pink and green arrows respectively.}
\label{Figure:Lattice}
\end{figure*}

Figure \ref{Figure:Lattice}(a,b) shows the optical and atomic force microscopic images of a typical graphene/h-BN sample that we have measured \cite{ZhangGY}. Palladium (Pd) or gold (Au) electrode was deposited to ground the sample in order to avoid charging during ARPES measurements. The height profile shows that graphene exhibits significant out-of-plane height variation $0.6\pm0.1 $ {\AA}.  The Moir$\acute{e}$ pattern period is extracted to be $15.6\pm0.4$ nm, which is larger than 14 nm derived from the 1.8\% lattice mismatch between graphene and h-BN (Fig.~1(c)). This suggests that the graphene is stretched by ${\approx}$ 0.2\% by the h-BN substrate, and the reduced lattice match leads to an expanded Moir$\acute{e}$ pattern. Figure \ref{Figure:Lattice}(d) shows ARPES data measured along the $\Gamma$-K direction. Dispersions from the $\pi$ and $\sigma$ bands of both graphene and h-BN are observed, and they overlap in most regions except the $\pi$ band near the K point. Here, characteristic linear dispersion from the Dirac cone of graphene is observed, while the top of the $\pi$ band from h-BN lies at -2.2 eV, in agreement with the large band gap of h-BN. The absence of graphene $\pi$ band splitting confirms that the sample is dominated by monolayer graphene. An expected result of the Moir$\acute{e}$ potential is the replicas of graphene Dirac cone at the same energy while translated by $\vec{G}_{s}$ from the graphene K point (first generation Dirac cones, see Fig.~1(e)).  The duplicated Dirac cones are obvious in both the Fermi surface map (Fig.~1(f)) and the dispersion images through the nearest SBZ centers (Fig.~1(g-i)). The average separation between the original Dirac band and its replicas is  $0.044\pm0.003$ ${\AA}^{-1}$, which is consistent with the observation of expanded Moir$\acute{e}$ pattern.

\begin{figure*}
\includegraphics[width=15.8 cm] {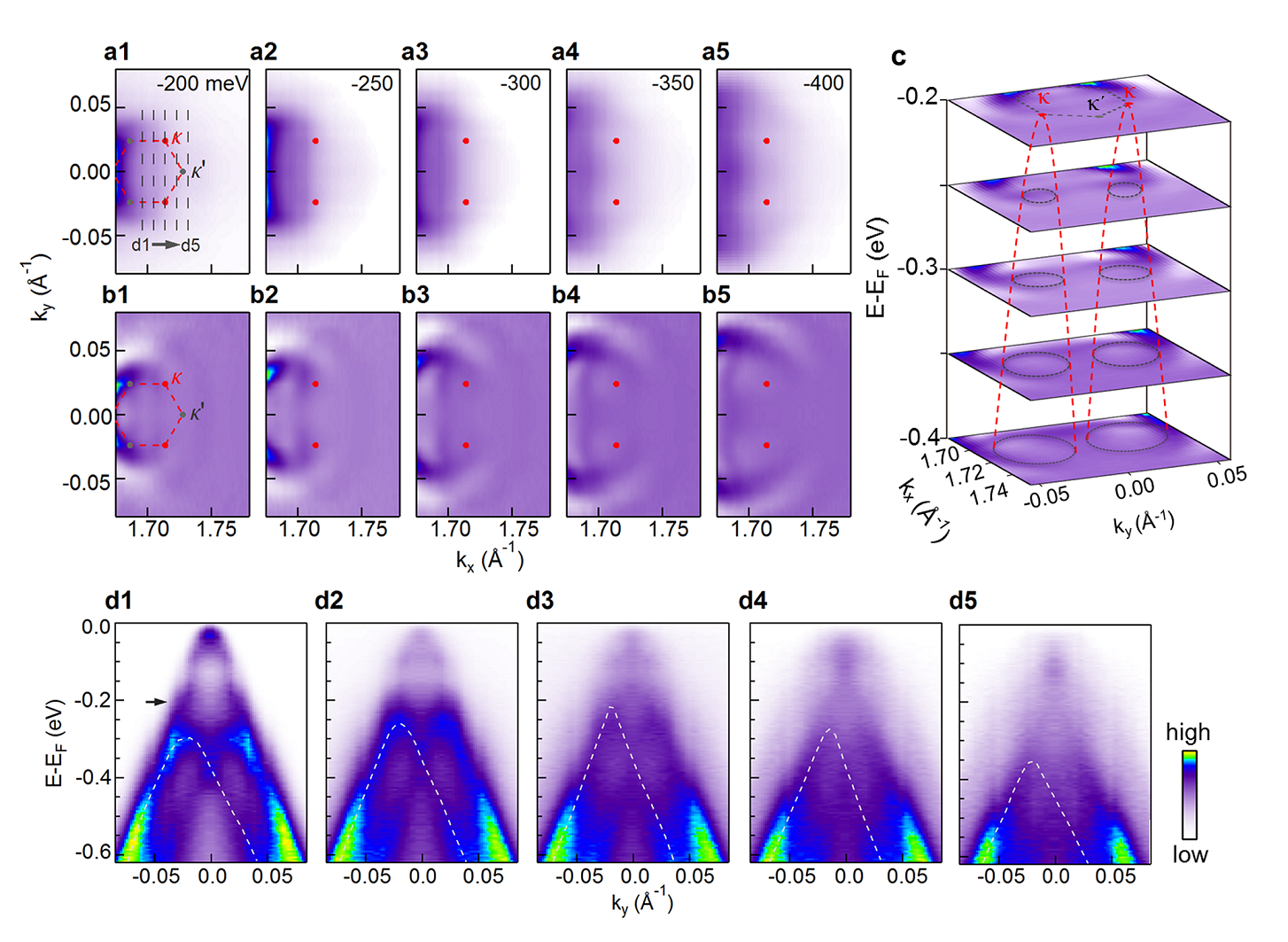}
\caption{{\bf Emergence of second generation Dirac cones.} (a1-a5) Constant energy maps at -200, -250, -300, -350 and -400 meV. (b1-b5) Constant energy maps of EDC curvature at -200, -250, -300, -350 and -400 meV. The Moir$\acute{e}$ SBZ is shown as broken red lines.  Red and grey dots indicate the inequivalent two sets of SBZ corners, $\kappa$ and $\kappa'$ . (c) Stacking of constant energy maps of EDC curvature to show the conical dispersion at two $\kappa$ points. (d1-d5) Dispersions along cuts d1-d5 shown in (a1). The white dashed lines are guides for the evolution of SDCs dispersions. The black arrow indicates the dispersion from the original Dirac cone. }
\label{Figure:Map}
\end{figure*}

We focus on data taken near the SBZ to search for signatures of SDCs away from E$_F$. Figure \ref{Figure:Map}(a1-a5) displays intensity maps taken at constant energies between -200 to -400 meV.  Corresponding curvature plots \cite{DingHRSI} are used to highlight dispersive bands in ARPES data (Fig.~2(b1-b5)).  At -250 meV, a small pocket is visible at the upper right and lower right corners of the SBZ (red dots, labeled as $\kappa$ in Fig.~2(a1)), and is especially clear at -350 meV and -400 meV. The size of the pockets grows when decreasing the energy, which is consistent with conical dispersions. Dispersion images (Fig.~2(d1-d5)) taken near these corners further support the existence of conical dispersions. When approaching the second generation Dirac points (SDPs) at $\kappa$ (cuts d1 through d3), the dispersion relations exhibit a rounded M-shape, with the  top  of these bands moving toward higher energies and reaching -0.21 eV at SDPs (cut d3). After passing the SDPs, the top of the valence bands move to lower energies (cuts d3 through d5), again in agreement with conical dispersion. We note that around the right corner of SBZ (labeled as $\kappa^\prime$), no conical dispersion is observed, suggesting that the two superlattice valleys $\kappa$ and $\kappa^\prime$ are inequivalent. Therefore, both the constant energy maps and the dispersion images presented above reveal directly that SDCs exist at the SBZ corners and only at one of the two superlattice valleys \cite{JeilNcomm}, which is in agreement with the Landau level degeneracy implied from previous quantum Hall effect measurements \cite {ZhangGY, GeimNaturePhys2014}. Such direct information is critical for determining the parameters used for describing the generic band structure of SDCs \cite{FalkoMiniband, JungPRB14}, and for further understanding other experimental results which probe the electronic structure indirectly.

\begin{figure*}
\includegraphics[width=10 cm] {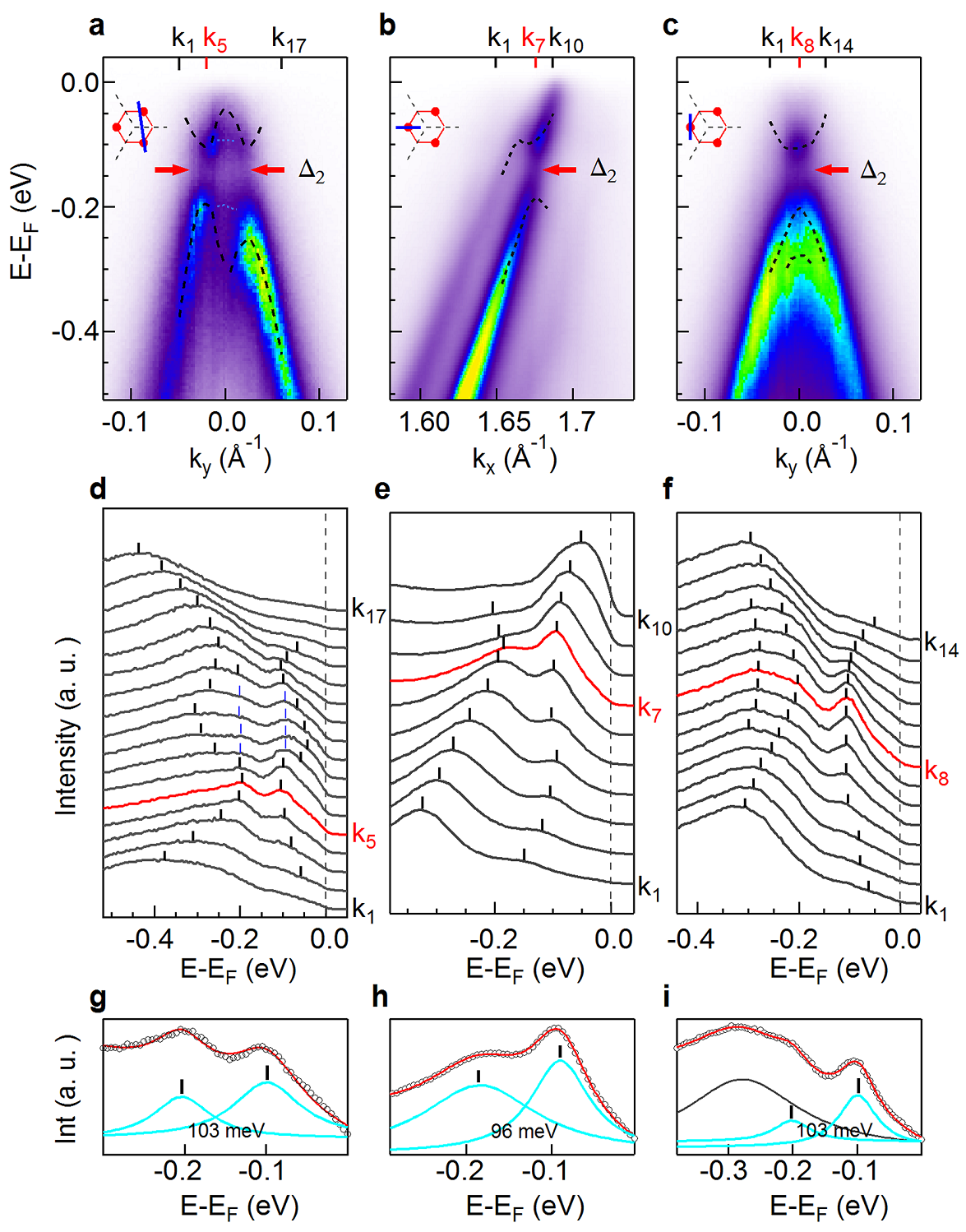}
\caption{{\bf Observation of gap opening at second generation Dirac points.} (a-c) ARPES data through SDPs along different directions. Black dashed lines are guides for the dispersion extracted from EDCs. Blue dotted lines indicate the non-dispersive states from impurity scattering. The insets are schematic drawings of the measurement geometry. The Graphene and superlattice brillouin zones are indicated by black dashed and red solid lines. The red dots represent the $\kappa$ points of SBZ. (d-f) EDCs between momentums indicated in (a-c). The EDCs across SDPs are highlighted by red lines. (g-i) Fitting results of the EDCs across SDPs with two (g-h) or three (i) Lorentzian peaks. The energy separation between the fitted Lorentzian peaks (ticks on cyan curves) indicates the gap size. }
\label{Figure:Gap1}
\end{figure*}

To better resolve the detailed dispersions of the SDCs, we show in Fig.~\ref{Figure:Gap1}(a) ARPES data taken near the $\kappa$ point of the SBZ. The conduction bands and valence bands show a rounded W- and M-shape respectively with a large energy separation and a suppression of intensity between them. This indicates the gap opening at the SDCs. The corresponding energy distribution curves (EDCs) are shown in Fig.~3(d). In addition to peaks from the conduction and valence bands, there are also non-dispersive peaks (marked by blue ticks) at the band extrema, suggesting that there are significant impurity scatterings at the gap edges.  From the extracted dispersions and the peak separation at the SDC (Figs.~\ref{Figure:Gap1}(g)), the gap at the SDCs is $\approx$ 100 meV. Direct observation of gapped SDC around the left corner of the SBZ (equivalent to the two $\kappa$ points discussed above) in the constant energy maps is difficult, since the intensity around this momentum region is dominated by the original graphene Dirac cone. However, the intensity suppression in EDCs (Figs.~\ref{Figure:Gap1}(e,f)) from data taken along two high symmetry directions $\Gamma$-K and $\Gamma$-M  (Figs.~\ref{Figure:Gap1}(b,c)) prove its existence with a similar gap (Figs.~\ref{Figure:Gap1}(h,i)). The observation of gapped SDCs at only one of the superlattice valleys $\kappa$ suggest that the inversion-asymmetric perturbation potential from the h-BN substrate plays an important role in the electronic structure of graphene/h-BN.

\begin{figure*}
\includegraphics[width=10 cm] {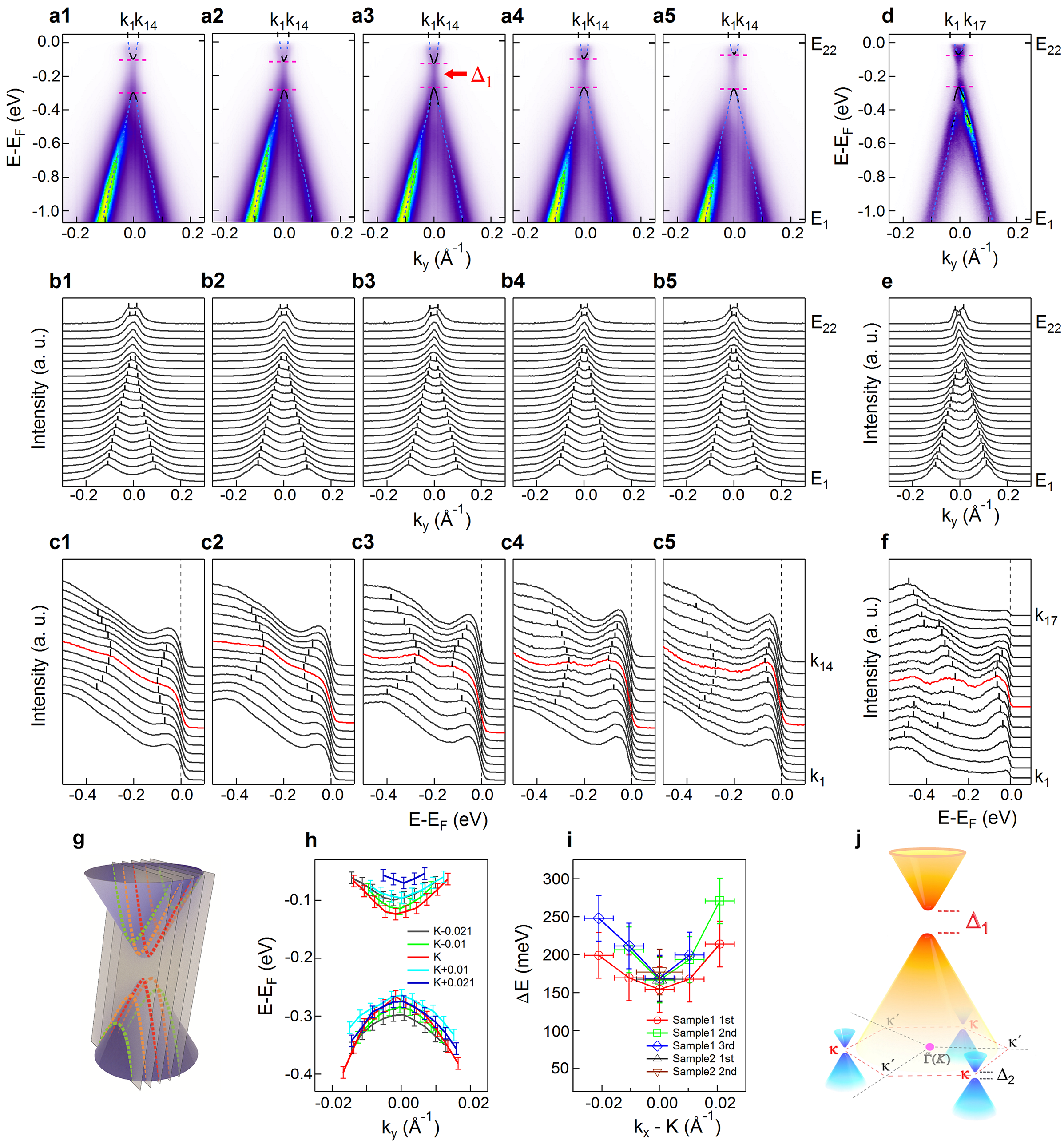}
\caption{{\bf Gap at the original Dirac point.}  (a1-a5) Band dispersions around the K point from sample 1 after 1st doping. The cuts are measured at k$_x$-K = -0.021, -0.01, 0, +0.01 and +0.021${\AA}^{-1}$ respectively. Blue dotted and black lines are guides for dispersions extracted from MDCs and EDCs respectively. The pink broken lines indicate the bottom of conduction band and the top of valence band. (b1-b5) Corresponding MDCs for (a1-a5) between energy $E_1$ and $E_{22}$. Black tick marks are guides for the peak positions. (c1-c5) Corresponding EDCs for (a1-a5) between momentum $k_1$ and $k_{14}$ to determine the dispersions around the Dirac point. The black tick marks are guides for dispersions from the Dirac cone. The EDCs at k$_y$=0 are highlighted by red color. (d-f) Same as (a3-c3) from sample 2 after 2nd doping. (g) Schematic drawing of the geometry for cuts a1-a5.  (h) Dispersions extracted from (c1-c5) to show the evolution of band dispersion when crossing the K point. The error bar is defined as the upper limit when the energy position is clearly offset the peak position. (i) Summary of the energy separation between the bottom of conduction band and the top of valence band for each cut measured on sample 1 and sample 2 at all doping levels. The minimum separation at the K point gives the gap size. The energy error bar is defined as the sum of the error bars of peak positions in (h). The momentum position of the K point is determined from the symmetry of a1-a5 and the error bar is set as half of the momentum step size. (j) Schematic drawing of the band structure in G/h-BN heterostructure, showing the gapped original Dirac point with gapped SDPs at three out of six corners of SBZ ($\kappa$ points). }
\label{Figure:Gap2}
\end{figure*}

To check whether there is a gap at the original Dirac cone, we dope the graphene/h-BN sample by depositing Rubidium to shift the Dirac point below E$_F$ so that it can be measured by ARPES. By shifting the Dirac point to -0.2 eV,  the conduction band becomes detectable. Figure \ref{Figure:Gap2}(a1-a5) shows a few cuts around the K point of graphene (see schematic drawing in (g)). Because photoemitted electrons from higher binding energy have lower kinetic energy and smaller in-plane momentum component, the cutting planes are slightly tilted.  The extracted dispersions from momentum distribution curves (MDCs, b1-b5) and EDCs (c1-c5) are overplotted as blue dotted and black solid curves in Fig.~4(a1-a5). Figure 4(h) focuses on the zoom-in dispersions near the Dirac point. The conduction and valence bands obviously do not touch each other when the cutting plane crosses the K point. Such behavior is an anambigous signature of an excitation  gap. Figure \ref{Figure:Gap2}(d-f) shows the data through the K point from sample 2. The dispersion is consistent with sample 1 while the intensity suppression from the gap at the SDCs is still clearly visible, likely because of the higher sample quality. Figure 4 (i) summarizes the energy separation between the conduction and valence bands.  From the minimum separation at the K point, the gap is extracted to be $160 \pm 30$ meV. The raw data for multiple samples at all doping levels can be found in the Supplementary Information.

The extracted gap size at the original Dirac cone from ARPES measurements is much large compared to previous measurements,  e.g. 15$\sim$30 meV from transport measurements in transferred graphene/h-BN \cite{GBNgap, NovoselovNphys2014} and $\approx$ 40 meV from magneto-optical spectroscopy on similar R-PECVD samples \cite{LiZQ}, as well as theoretical predictions. Here we discuss a few reasons and the implications.  First of all, since the gap is extracted from the band edges in ARPES and from the density of states in other measurements, gaps probed by other techniques can be much smaller than by ARPES since the density of states rarely drops to zero abruptly at the band edges.  Impurities can also contribute to in-gap states and lead to a smaller gap size. In bilayer graphene, it has been suggested that the gap extracted from transport or STM measurements can be underestimated due to additional conductive channels by defects and charge impurities \cite{StroscioNaturePhys2011}, and this may also affect the gap size measured in graphene/h-BN.  Secondly, although the maximum gap at the equilibrium layer separation for perfectly lattice matched heterostructure is predicted to be $\approx$ 50 meV  \cite{DFTgap}, the gap size increases sharply upon reducing the separation between graphene and h-BN layers \cite{DFTgap}. A band gap opening at the graphene Dirac cone and SDCs can be induced by an inversion asymmetric mass term in the perturbation potential  \cite{FalkoMiniband, MoonPRB14, JeilNcomm}, and the mass term can vary by orders of magnitude depending on the assumptions used \cite{DFTgap, MillerPRB12, LevitovPRL13}.  Shortening of the layer separation between graphene and h-BN by 0.2 \AA  ~around the equilibrium distance can result in almost a threefold increase of the local mass term \cite{DFTgap}.  Thirdly,  the substantial out-of-plane height variation of 0.6$\AA$ and 0.2$\%$ in-plane strain revealed by AFM measurements suggest large enough variations in the local mass term and can produce a large average gap \cite{Bokdam,DFTgap}. Therefore, the large gap size suggests that our epitaxial grown graphene/h-BN samples have much stronger short-range interlayer interaction than weak van der Waals interaction as commonly believed in ideally flat structures. Although a definitive explanation of the gap opening still requires more theoretical and experimental investigations, our work reports the intriguing electronic structure in a model van der Waals heterostructure and highlights the important role of the inversion symmetry breaking perturbation potential and interfacial atomic structure in the physics of graphene/h-BN heterostructure.

{\bf Methods}

{\bf Sample preparation.} Graphene samples were directly grown on h-BN substrates by the epitaxial method as specified in ref.\cite {ZhangGY}. As-grown samples were characterized by tapping mode AFM (MultiMode IIId, Veeco Instruments) at room temperature in ambient atmosphere. We used freshly cleaved mica as shadow masks for metal electrode deposition. Using the micromanipulator mounted on an optical microscope, the freshly cleaved mica flakes were accurately transferred on the substrates, covering one side of the substrates with most area of target graphene/h-BN samples. The contact metal ($\approx$ 40 nm Pd or Au) was deposited on the non-mica-covered area with a small part of target graphene/h-BN samples. The samples were then annealed at 200$^\circ$C, after removing the mica flakes.

{\bf ARPES measurement.} ARPES measurements have been performed using ARPES instruments based at the synchrotron-radiation light source at beamline 12.0.1 of the Advanced Light Source at Lawrence Berkeley National Laboratory (LBNL). The data were recorded by Scienta R3000 at BL 12.0.1 with 50 and 60 eV photon energy. The energy and angle resolution is better than 30 meV and 0.2$^\circ$ respectively. Before measurement, the samples were annealed at 200$^\circ$C for 2-3 hours until clean surfaces were exposed. All measurements were performed between 10-20 K and under vacuum better than $5 \times 10^{-11}$ Torr. The Rubidium deposition was achieved by heating up SAES commercial dispenser in situ.

~\\

{\bf Acknowledgements}
We thank V.I. Fal'ko for the useful discussions. This work is supported by the National Natural Science Foundation of China (Grant No.~11274191, 11334006, and 11427903), Ministry of Science and Technology of China (Grant No.~2015CB921001, 2016YFA0301004) and Tsinghua University Initiative Scientific Research Program (2012Z02285). E.Y.W. thanks support from the Advanced Light Source doctoral fellowship program. The Advanced Light Source is supported by the Director, Office of Science, Office of Basic Energy Sciences, of the U.S. Department of Energy under Contract No. DE-AC02-05CH11231.

{\bf Author Contributions}
S.Z. designed the research project. E.W., S.D., W.Y., M.Y., G.W., K.D., A.V.F. and S.Z. performed the ARPES measurements and analyzed the ARPES data.  X.L., S.W., G.C., G.Z and Y. Z. prepared the graphene samples. J.J. discussed the data. E.W. and S.Z. wrote the manuscript, and all authors commented on the manuscript.

{\bf Competing financial interests}
The authors declare no competing financial interests.

\newpage
\begin{center}
{\large Supplementary Information}
\end{center}

\section{ Raw data at different doping levels of different samples}

\begin{figure*}
\includegraphics[width=16.8 cm] {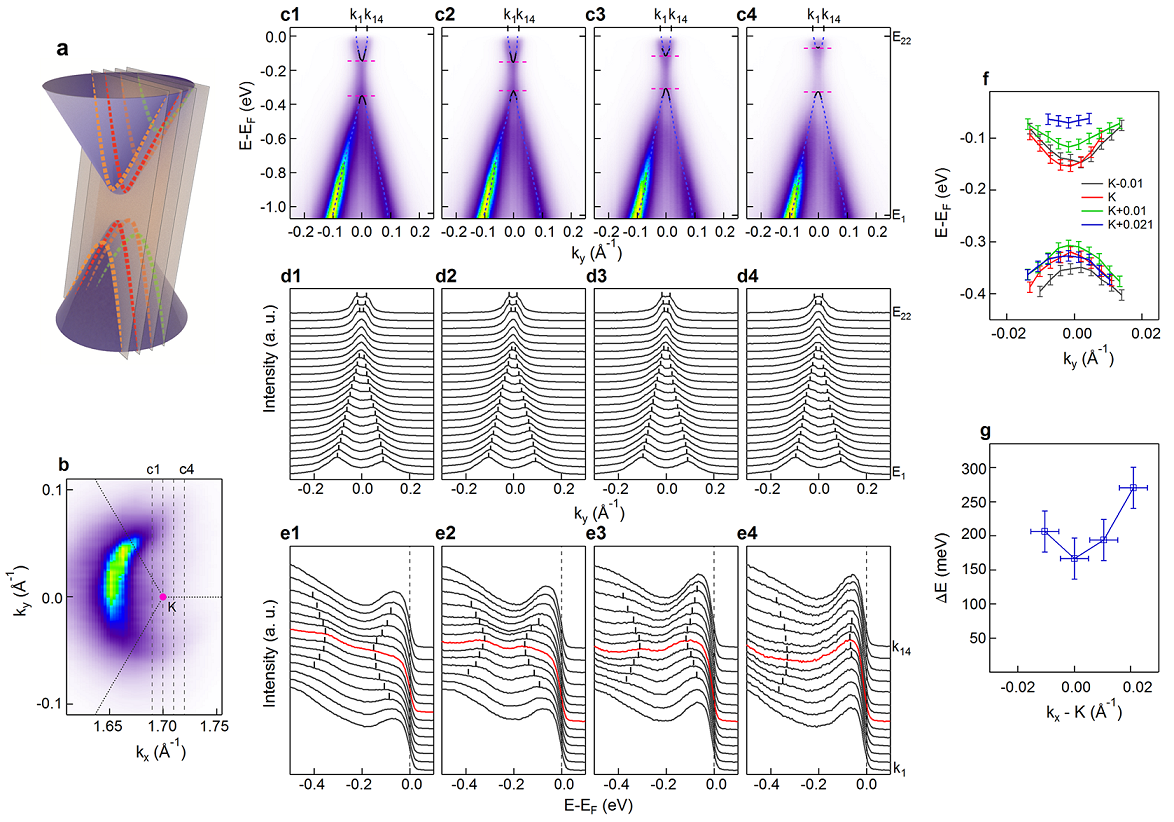}
\renewcommand{\thefigure}{S\arabic{figure}}
\caption{{\bf Band dispersions near the original Dirac point for sample 1 after 2nd doping.} (a) Schematic drawing of the geometry when taking the cuts c1-c4. (b) Constant energy map at E$_D$-450 meV  to show the positions of cuts c1-c4. E$_D$ is the energy position of Dirac point. (c1-c4) Band dispersions around the K point from sample 1 after 1st doping. Blue dotted and black lines are guides for dispersions extracted from MDCs and EDCs respectively. The pink broken lines indicate the bottom of conduction band and the top of valence band. (d1-d4) Corresponding MDCs for (c1-c4) between energy $E_1$ and $E_{22}$. The black tick marks are guides for the peak positions. (e1-e4) Corresponding EDCs for (c1-c4) between momentum $k_1$ and $k_{14}$ to determine the dispersions around Dirac point. The black tick marks are guides for dispersions from the Dirac cone. The EDCs at k$_y$=0 are highlighted by red color. (f) The dispersions extracted from (e1-e4) to show the evolution of band dispersion when crossing the K point. The error bar is defined as the upper limit when the energy position is clearly offset the peak position. (g) Summary of the energy separation between the bottom of conduction band and the top of valence band for each cut. The energy error bar is defined as the sum of the error bars of peak positions in (f). The momentum position of the K point is determined from the symmetry of c1-c4 and the error bar is set as half of the momentum step size.}
\label{Figure:SupS1}
\end{figure*}

\begin{figure*}
\includegraphics[width=16.8 cm] {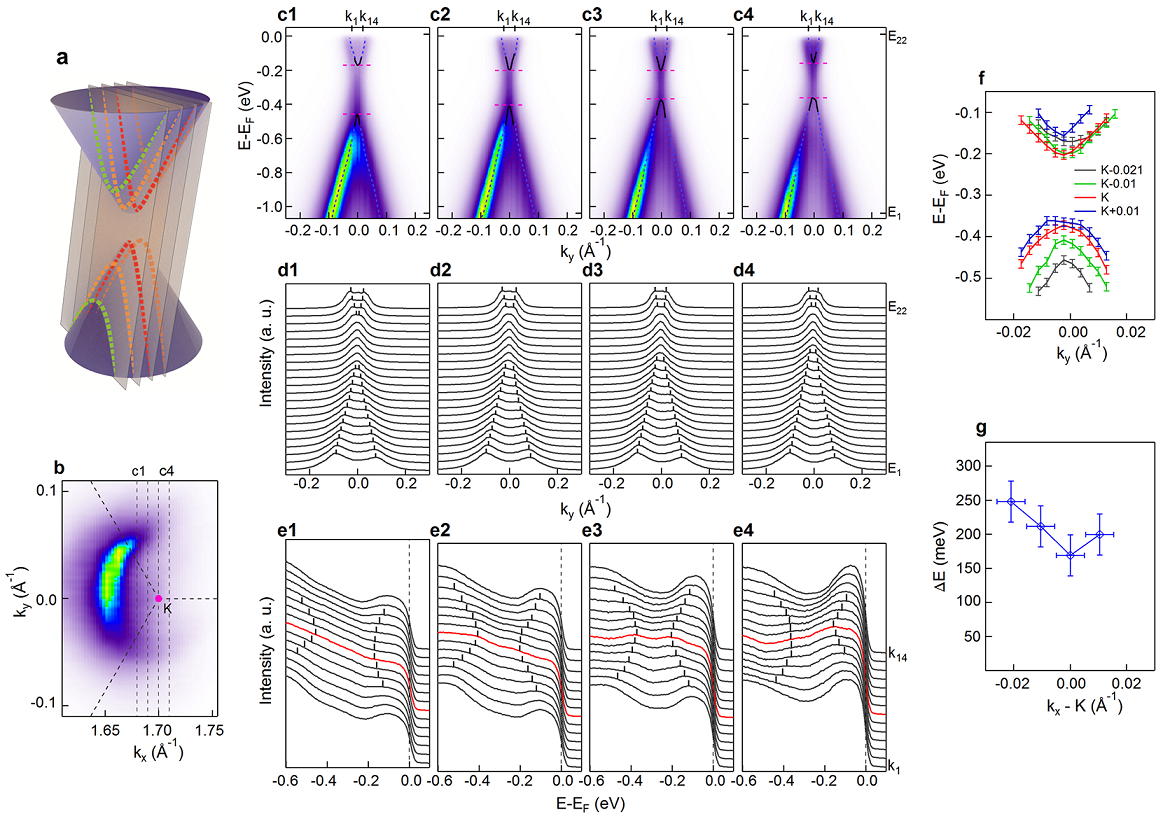}
\renewcommand{\thefigure}{S\arabic{figure}}
\caption{{\bf Band dispersion near the original Dirac point for sample 1 after 3rd doping.} (a) Schematic drawing of the geometry when taking the cuts c1-c4. (b) Constant energy map at E$_D$-450 meV to show the positions of cuts c1-c4. E$_D$ is the energy position of Dirac point. (c1-c4) Band dispersions around the K point from sample 1 after 1st doping. Blue dotted and black lines are guides for dispersions extracted from MDCs and EDCs respectively. The pink broken lines indicate the bottom of conduction band and the top of valence band. (d1-d4) Corresponding MDCs for (c1-c4) between energy $E_1$ and $E_{22}$. The black tick marks are guides for the peak positions. (e1-e4) Corresponding EDCs for (c1-c4) between momentum $k_1$ and $k_{14}$ to determine the dispersions around Dirac point. The black tick marks are guides for dispersions from the Dirac cone. The EDCs at k$_y$=0 are highlighted by red color. (f) The dispersions extracted from (e1-e4) to show the evolution of band dispersion when crossing the K point. The error bar is defined as the upper limit when the energy position is clearly offset the peak position. (g) Summary of the energy separation between the bottom of conduction band and the top of valence band for each cut. The energy error bar is defined as the sum of the error bars of peak positions in (f). The momentum position of the K point is determined from the symmetry of c1-c4 and the error bar is set as half of the momentum step size.}

\label{Figure:SupS2}
\end{figure*}

\begin{figure*}
\includegraphics[width=10 cm] {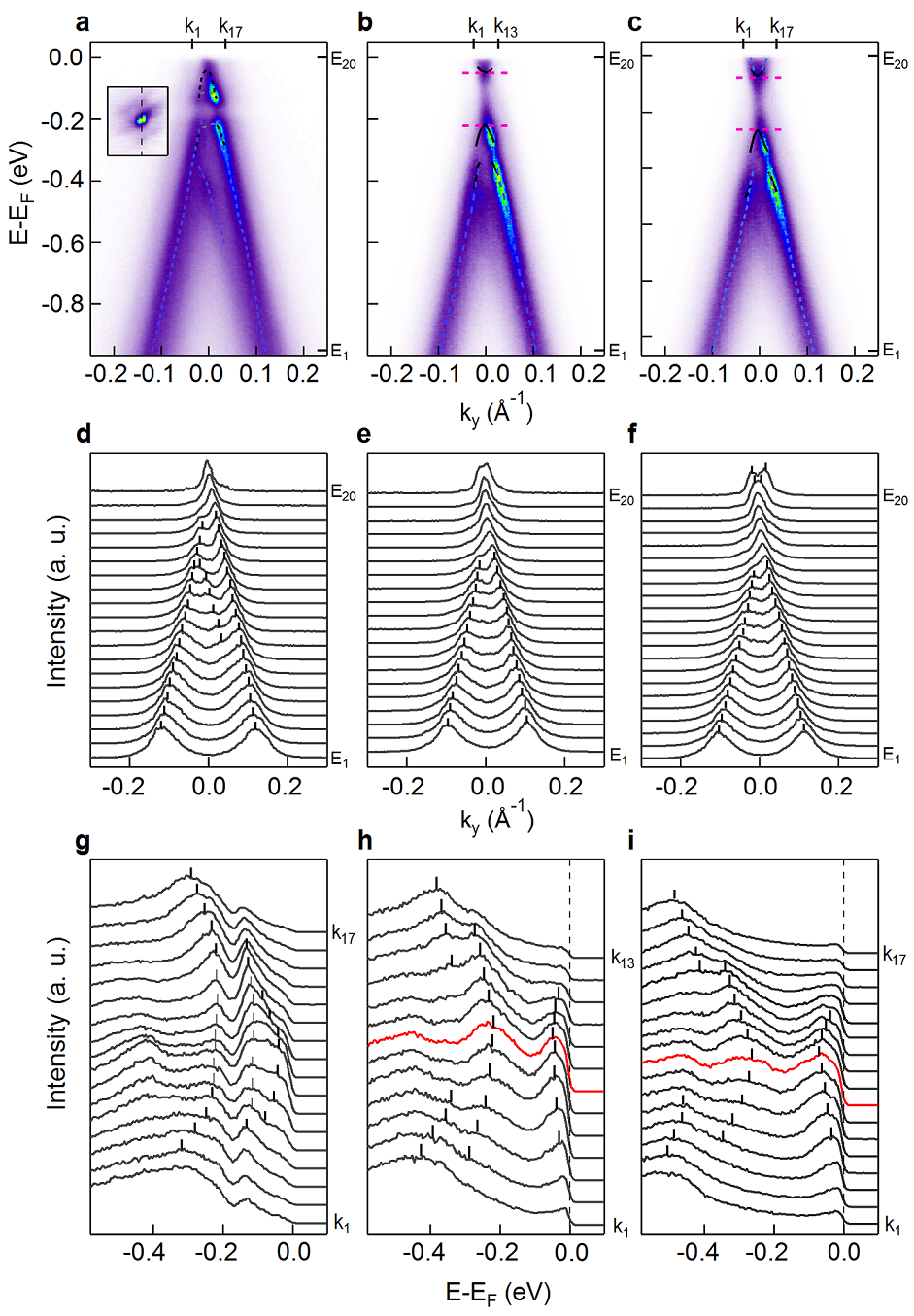}
\renewcommand{\thefigure}{S\arabic{figure}}
\caption{{\bf Band dispersions at the original Dirac point for sample 2.} (a) Pristine band dispersions at Dirac point for sample 2. Inset shows the constant energy map at Fermi energy. (b) Band dispersions after 1st doping. (c) Band dispersions after 2nd doping, same as in Fig.~4(f).  In (a-c), blue dotted and black lines are guides for dispersions extracted from MDCs and EDCs respectively. The pink broken lines indicate the bottom of conduction band and the top of valence band. (d-f) Corresponding MDCs for (a-c) between energy $E_1$ and $E_{20}$. The black tick marks are guides for the peak positions. (g-i) Corresponding EDCs for (a-c) between momentum $k_1$ and $k_{17}$ ($k_{13}$ for (h)) to determine the dispersions around Dirac point. The black tick marks are guides for dispersions. The EDCs at k$_y$=0 are highlighted by red color.}

\label{Figure:SupS3}
\end{figure*}

For determining the gap size at the original Dirac point, in order to avoid the artifact due to derivation from K point, we first took maps around K point by $0.2^\circ$ step to determine the accurate position of Dirac point. The errorbar of the Dirac point position is within $0.1^\circ$ in this case, which converts to 0.006 ${\AA}^{-1}$ and 0.0066 ${\AA}^{-1}$ for 50 and 60 eV photo energy respectively. For sample 1, we took several cuts around the Dirac point after each doping (Fig.~4, Fig.~\ref{Figure:SupS1}, \ref{Figure:SupS2}). The panels (c1-c4) in Fig.~\ref{Figure:SupS1} and Fig.~\ref{Figure:SupS2} show the band dispersions near the Dirac point for sample1 after 2nd and 3rd doping. The extracted dispersions from peaks in the momentum distribution curves (MDCs, panels (d1-d4)) where the band dispersions are linear and the energy distribution curves (EDCs, panels (e1-e4)) near the Dirac point are overplotted as  blue dotted and black solid lines in panels (c1-c4), and the zoom-in dispersions near the Dirac point are shown in panel (f). The bottom of conduction band and the top of valence band do not touch each other when the cutting plane crosses the K point, signalling the gap opening at the original Dirac point, consistent with the conclusions in main text. The evolution of the minimum separation between the valence and conduction bands measured at each cut across the original Dirac cone is shown in panel(g). For sample 2, we deposited Rubidium on sample surface in situ after moving sample to the angle corresponding to the K point. The suppression of intensity at the original Dirac point is also visible after shifting Dirac point below Fermi energy which signals gap opening (Fig.~\ref{Figure:SupS3}). The evolution of the minimum separation between the valence and conduction bands measured at each cut around the original Dirac cone from sample 1 and sample 2 at different doping levels is summarized in Fig.~4(i).

\end{document}